\documentclass[aps,preprint,prc,showpacs,nofootinbib,superscriptaddress]{revtex4-1}

\usepackage[colorlinks]{hyperref}

\usepackage{graphicx}
\usepackage{amsmath,amssymb}
\usepackage{wasysym} 

\newcommand{\beq}{\begin{equation}}
\newcommand{\eeq}{\end{equation}}
\newcommand{\bes}{\begin{split}}
\newcommand{\ees}{\begin{split}}
\newcommand{\bea}{\begin{eqnarray}}
\newcommand{\eea}{\end{eqnarray}} 
\newcommand{\nn}{\nonumber \\ }
\newcommand{\bfi}{\begin{figure}}
\newcommand{\efi}{\end{figure}}

\begin{document}

\title{Three-body systems in physics of cold atoms and halo nuclei}

\author{Chen Ji}
\email{ji@ectstar.eu}
\affiliation{ECT*, Villa Tambosi, 38123 Villazzano (Trento), Italy}
\affiliation{INFN-TIFPA, Trento Institute for Fundamental Physics and Applications, Trento, Italy}
\affiliation{TRIUMF, 4004 Wesbrook Mall, Vancouver, BC V6T 2A3, Canada}

\date{\today}

\begin{abstract}
Few-body systems, such as cold atoms and halo nuclei, share universal features at low energies, which are insensitive to the underlying inter-particle interactions at short ranges. These low-energy properties can be investigated in the framework of effective field theory with two-body and three-body contact interactions. I review the effective-field-theory studies of universal physics in three-body systems, focusing on the application in cold atoms and halo nuclei. 
\end{abstract}



\maketitle

\section{Introduction}
\label{chap:1-intro}
Although physics in cold atoms and halo nuclei are driven by interactions at very different physical scales, these systems share common features in their respective low energy regimes. 
Universal behavior occurs when a system satisfies a separation of a large length scale and a small one. The large length scale is characterized by the scattering length $a$, which determines the total cross section of the two-body s-wave scattering at zero energy by $\sigma = 4\pi a^2$.\footnote{$\sigma =2\pi a^2$ for identical fermions.}
The small length scale is represented by the range of two-body interactions $\ell$.
In the limit $|a|\gg\ell$, physics at the scale of $a$ is disentangled from physics at the scale of $\ell$, and is therefore insensitive to the details of the short-range interactions.

An example of three-body universality is Efimov physics. In systems with three identical bosons, Vitaly Efimov predicted that, in the unitary limit $|a|\rightarrow\infty$, an infinite number of three-body bound states ({\it trimers}) emerge and accumulate at zero energy~\cite{Efimov70}. These trimers have a geometric spectrum that satisfies a discrete scaling symmetry.
This spectrum behavior, together with many other few-body features satisfying the discrete scaling symmetry in the limit $|a|\gg \ell$, are often called ``the Efimov effect''. Evidence of the Efimov effect was found in the recombination processes in ultracold atomic gases, such as $^{133}$Cs~\cite{Kraemer:2006,knoop:2008,Zenesini2014,Huang2014}, $^{7}$Li~\cite{Gross:2009,Gross:2010,Pollack:2009}, $^{39}$K~\cite{Zaccanti2009}, and $^{85}$Rb~\cite{Wild2012}.
In these experiments, the atom-atom scattering length $a$ is tuned through an external magnetic field to arbitrarily large values near Feshbach resonances~\cite{Chin:2008}, where free atoms form shallow dimers (two-atom bound states) or trimers. The atomic recombination rates are measured as a function of $a$. By tuning the magnetic field, critical features such as recombination minima and resonances occur at different values of $a$. The discrete scaling symmetry has been observed in the critical recombination phenomena, which are labeled by the values of $a$.

Universality also exists in molecular clusters of helium atoms. As observed by Luo {\it et al.}~\cite{luo1993}, two $^4$He atoms form a shallow dimer. The atom-atom scattering length is $\sim 100$ \AA, about $20$ times the range of the van der Waals potential~\cite{Naidon:2011,Cencek:2012}. The $^4$He trimer has been calculated using various realistic potential models~\cite{Roudnev2000, Roudnev2003, Motovilov:1999iz, Barletta:2001za}, which indicated the existence of two consecutive (ground and excited) trimer states. The ground state was observed two decades ago~\cite{Schollkopf:1994}, but it is only until recently that the excited-state trimer has been observed using Coulomb explosion imaging techniques~\cite{Kunitski:2015}.

In nuclear systems, the nucleon-nucleon s-wave scattering length is 3 times the range of nuclear forces (the inverse pion mass) in the spin-triplet channel, and is 15 times in the singlet channel~\cite{Mathelitsch:1984hq}. This separation of scales yields universal properties in few-nucleon systems. For example, the calculated values of the triton binding energy and the spin-doublet neutron-deuteron scattering length obey a linear correlation, which does not depend on the nucleon-nucleon potential models or potential parameterizations. This linear correlation is well known as the Phillips line~\cite{Phillips68}.

Another candidate for investigating few-body universal physics is the halo nucleus~\cite{Zhukov-93,Tanihata1996,Jensen-04}, {\it i.e.} a nucleus that contains one or several nucleons loosely attached to a tightly bound nuclear core. The valence nucleons can move far away from the core, thus forming a halo that substantially extends the spatial distribution of the nucleus. The scale separation between the shallow valence-nucleon separation energy and the deep core excitation energy allows connecting the clustering mechanism in halo nuclei with universal features.

One successful approach to describe universal physics in few-body systems is an effective field theory (EFT). This theory utilizes the separation of scales and integrates out the short-range dynamics beyond the EFT description. The short-range effects to low-energy physics are embedded into a series of two- and three-body effective contact interactions, which are constructed based on a systematic expansion of the ratio between two momentum scales, $Q/\Lambda$. The low momentum $Q\sim 1/a$ denotes the typical momentum of particles in the system, and the high momentum $\Lambda\sim1/\ell$ quantifies when the EFT breaks down. The coupling constants of the counterterms are determined from low-energy observables. The resulting EFT with contact interactions is known as the pionless EFT~\cite{Kaplan:1996xu,Kaplan:1998tg, vanKolck:1998bw,Birse:1998dk,vanKolck:1999mw,Beane:2000fx,Bedaque:2002mn,Rho:2002sh} in nuclear physics. It has also been applied to cold atomic and halo physics, and is often dubbed respectively as short-range EFT ({\it e.g.} in Refs.~\cite{Platter:2008cx,Platter:2009gz,Ji:2011qg}) and halo EFT ({\it e.g.} in Refs.~\cite{Bertulani:2002sz,Canham:2008jd,Higa:2008rx}). I will refer hereafter effective field theories with contact interactions simply as ``EFT''.

Detailed reviews of Efimov signatures in cold atomic physics~\cite{Braaten:2004rn} and nuclear/particle physics~\cite{Hammer:2010kp} already exist in the literature.
In this review, I will discuss the study of three-body universal physics using EFT approaches, focusing on the description of cold atomic and halo nuclear systems. Based on the systematic expansion in $Q/\Lambda$, we discuss the leading-order EFT predictions, the extension to various higher-order effects and other contributions.


\section{Universal physics in cold atoms}

\subsection{EFT for three identical bosons}
\label{sec:2-2-1-lag}

The system of three identical bosons interacting with short-range potentials has been studied by Bedaque {\it et al.}~\cite{Bedaque:1998kg} using EFT in the limit $|a|\gg\ell$. An effective Lagrangian is constructed as a series of two- and three-body contact interactions:
\begin{eqnarray}
\label{eq:Lagrangian}
\ensuremath{\mathcal{L}}&=&
\psi^\dagger\left(i\partial_0 +
  \frac{\nabla^2}{2m}\right)\psi +\sigma T^\dagger\left(i\partial_0 +
  \frac{\nabla^2}{4m}-\Delta \right)T
\nn
&&-\frac{g}{\sqrt{2}}\left(T^\dagger
  \psi\psi+\textrm{h.c}\right)+hT^\dagger T \psi^\dagger \psi +
\ldots~,
\end{eqnarray}
where $\psi$ and $T$ represents respectively the single boson field and the auxiliary dimer field. $\Delta$ indicates the bare mass of the dimer and $g$ ($h$) is the two-body (three-body) coupling constant. $\sigma=\pm 1$ with the sign determined from the effective range $r_0$. The ellipsis represents higher-order two- and three-body interactions in the EFT expansion, which contains either more derivatives or more fields. 

Using renormalization-group methods, several groups~\cite{Birse:1998dk,Kaplan:1996xu,Kaplan:1998tg,vanKolck:1998bw} 
showed that the EFT expansion is equivalent to the effective-range expansion for a scattering momentum $k<1/\ell$. The later expands the two-body low-energy s-wave phase shift in powers of $k^2$ by~\cite{Bethe:1949}
\begin{equation}
\label{eq:range-expansion}
k\cot \delta_0 = -\frac{1}{a} + \frac{1}{2}r_0 k^2+\cdots~,
\end{equation}
where $r_0$ denotes the effective range and is of order $\ell$. The ellipsis represents the shape-parameter and higher--effective-range terms of order $k^4$ and above. The EFT description of a system satisfying $|a|\gg\ell$ instead expands the two-body s-wave scattering amplitude, $\propto (k\cot \delta_0-ik)^{-1}$, around the singularity pole. This expansion is equivalent to Eq.~\eqref{eq:range-expansion} provided that $k<1/\ell$. 
The leading order (LO) EFT corresponds to parameterizing physical quantities in terms of $a$. Since terms of order $k^4$ do not enter EFT calculations until three orders beyond LO, the inclusion of range effects can describe the large--scattering-length physics at least up to a next-to-next-to-leading-order (N$^2$LO) accuracy.
Corrections from finite-range effects can be evaluated perturbatively in the $r_0/a$ expansion.
Therefore, the four-momentum dimer propagator $\mathcal{D}(p)$ is expanded in powers of $r_0$ as $\mathcal{D}(p) \equiv \sum_{n=0}\mathcal{D}^{(n)}(p)$, with the $n$th-order piece defined by
\begin{equation}
\label{eq:dimer-expan}
\mathcal{D}^{(n)}(p)=\frac{4\pi}{m} \left(\frac{r_0}{2}\right)^n
\frac{\left(\gamma+\sqrt{-mp_0+{\bf p}^2/4}\right)^n}{-\gamma+\sqrt{-mp_0+{\bf p
}^2/4-i\epsilon}}~,
\end{equation}
where $\gamma$ is the two-body binding momentum obeying $1/a=\gamma-r_0\gamma^2/2$.

\begin{figure}[ht]
\centerline{\includegraphics*[width=1.\linewidth,angle=0,clip=true]{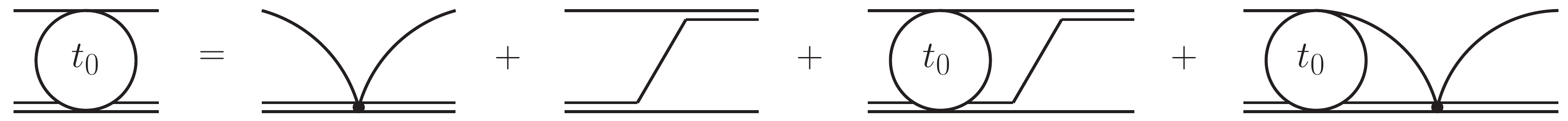}}
\caption{The LO amplitude $t_0$ for atom-dimer scattering in the integral equation: the double line and the round dot indicate the LO dimer propagator $\mathcal{D}^{(0)}(p)$ and three-body coupling.}
\label{pic:t0}
\end{figure}

The three-body problem can be studied by calculating the atom-dimer scattering amplitude. As illustrated in Fig.~\ref{pic:t0}, contributing diagrams to the LO amplitude $t_0$, which arise from the exchange of an atom between a dimer and an atom (proportional to $g^2$) and the atom-dimer contact interaction (proportional to $h$), need to be iterated due to the non-perturbative feature of large--scattering-length physics. 

$t_0$ is solved in the modified Skorniakov-Ter-Martirosian (STM) equation~\cite{STM57, Bedaque:1998kg}, which is equivalent to a non-relativistic Faddeev equation~\cite{Faddeev:1960su} with two- and three-body contact interactions. The s-wave projection yields
\begin{align}
\label{eq:t0}
t_0(k,p;E)=&M(k,p;E)+\frac{2}{\pi}\int_0^\Lambda dq
\frac{q^2 M(q,p;E) }{-1/a+\sqrt{3q^3/4-mE-i\epsilon}} t_0(k,q;E)
~,
\end{align}
where the kernel function is defined by
\begin{equation}
\label{eq:kernel-M}
 M(q,p;E)=\frac{1}{qp} \log\left(
  \frac{q^2+p^2+qp-mE}{q^2+p^2-qp-mE}\right)
+\frac{2H_0(\Lambda)}{\Lambda^2}~. 
\end{equation}
Here the three-body coupling $H_0 = \Lambda^2 h/2mg^2$, with $\Lambda$ denoting the regulation cutoff of the integral equation. $H_0$ is tuned to fit one three-body observable to ensure a cutoff independent result.

\begin{figure}[ht]
\centerline{\includegraphics*[width=0.6\linewidth,angle=0]{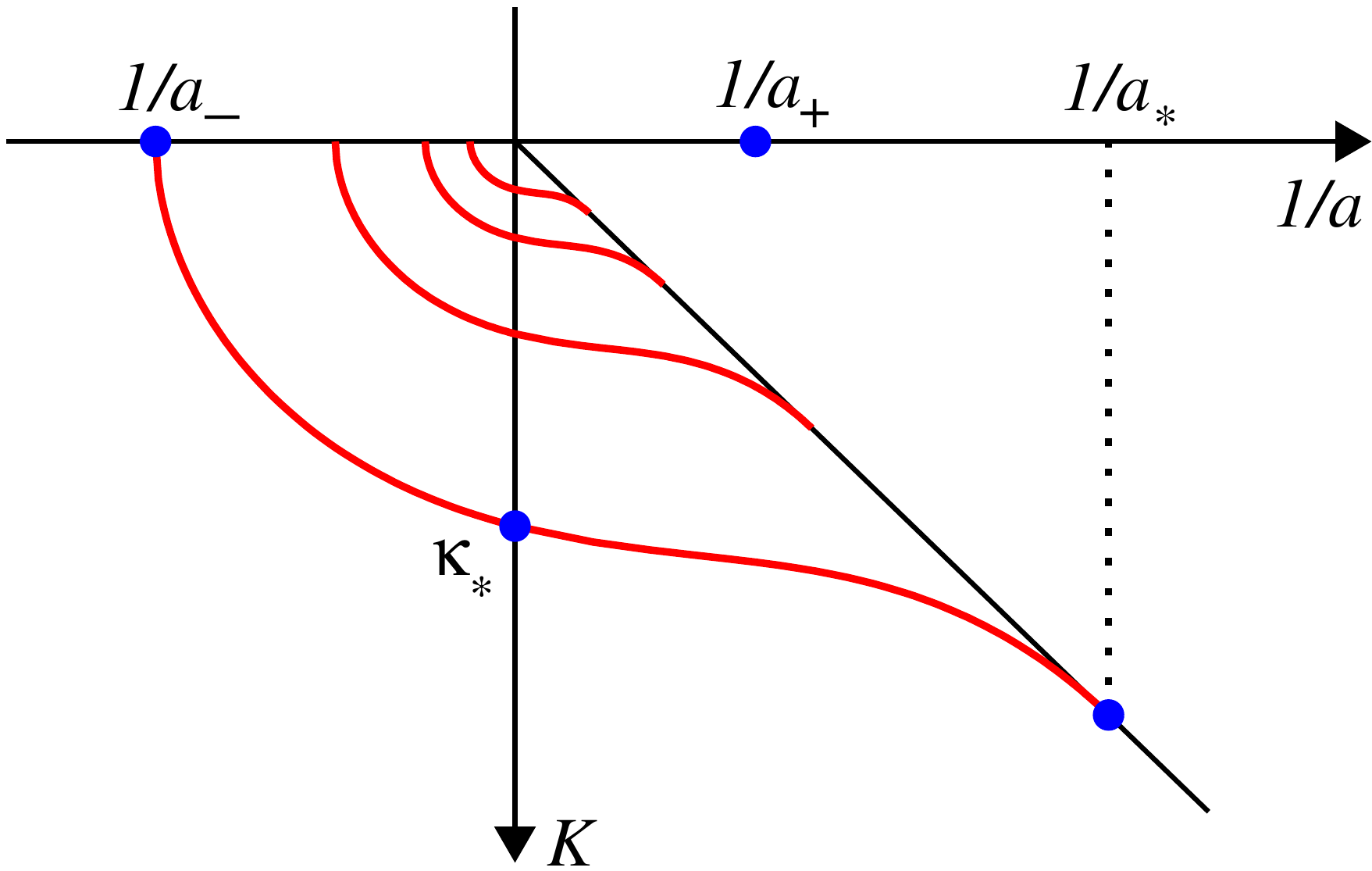}}
\caption
{The binding momentum $K$ of Efimov trimers is plotted as a function of $1/a$. The discrete scaling constant is artificially adjusted for ease of illustration.}
\label{pic:efimov}
\end{figure}

The bound-state case is solved in Eq.~\eqref{eq:t0} without the inhomogeneous term. The binding energies $B_0$ of Efimov trimers are manifested as poles in $t_0$.
In Fig.~\ref{pic:efimov}, the binding momentum $K=\sqrt{mB_0}$ is plotted as a function of $1/a$, where several critical phenomena appear along the spectrum curve: 
(a) the binding momentum $\kappa_*$ in the unitary limit $a \rightarrow \infty$,
(b) the scattering length  $a_{-}<0$ where free atoms form the Efimov trimer at zero energy,
(c) the scattering length $a_{*}>0$ at which the Efimov trimer dissociates into an atom and a dimer,
and (d) $a_{+}>0$ where the recombination reaches a local minimum. These Efimov features are related by universal numbers predicted from the LO EFT:
\begin{equation}
\label{eq:a-kappa*}
a_{i} \kappa_* = \theta_i,
\end{equation}
where $\theta_-=-1.50763$, $\theta_+= 0.316473$~\cite{Gogolin2008}, and $\theta_* = 0.0707645$~\cite{Braaten:2004rn}.

Fig.~\ref{pic:efimov} also displays a series of Efimov trimers obeying a discrete scaling symmetry. For example, the binding momentum $\kappa_{*,n}$ of the $n$th Efimov trimer in the unitary limit is related to $\kappa_*$ in the $0$th branch by 
\begin{equation}
\label{eq:Buni}
\kappa_{n} =  \lambda^{-n} \kappa_*, 
\end{equation}
with $\lambda=22.694$ denoting the scaling constant. Similarly, other Efimov features encoding the same critical phenomena also obey the scaling symmetry:
\begin{equation}
\label{eq:ani}
a_{i,n} = \lambda^{n} a_{i}~. 
\end{equation}

\begin{figure}[ht]
\centerline{\includegraphics*[width=0.6\linewidth,angle=0,clip=true]{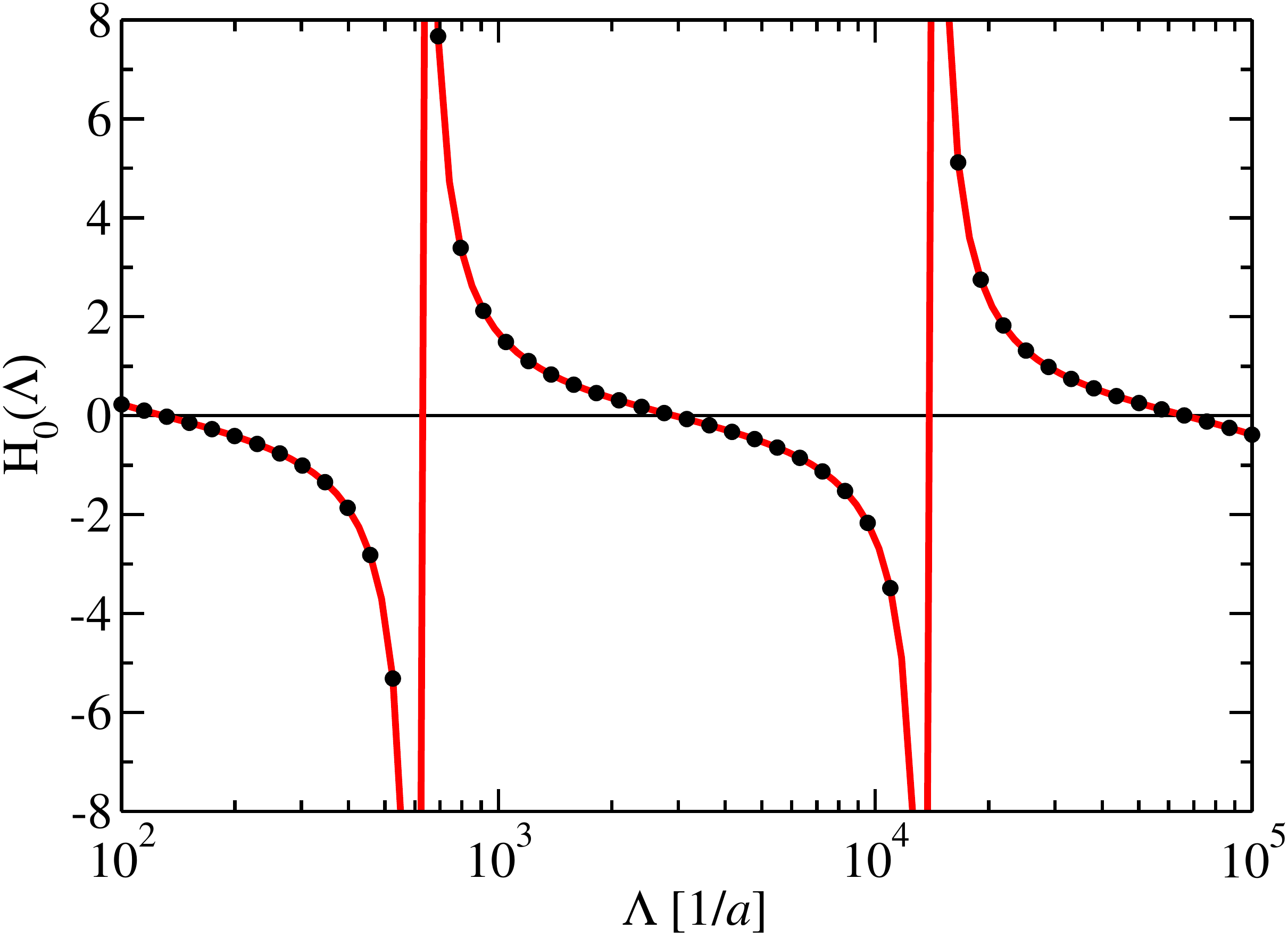}}
\caption{The LO three-body running coupling $H_0$ as a function of the cutoff $\Lambda$ at a fixed renormalization condition. The dots and solid line represent respectively the numerical results and the analytic expression given in Eq.\eqref{eq:h0}.}
\label{pic:HLambda}
\end{figure}

The discrete scale invariance can be explained as a result of  a limit cycle under the renormalization-group flow~\cite{PhysRevD.3.1818}. Several such analyses have been discussed in the EFT framework~\cite{Barford:2004fz,Braaten:2004pg,Griesshammer:2005ga,Hammer:2011kg}.
The three-body running coupling $H_0(\Lambda)$ has an analytic expression in the zero-range limit~\cite{Bedaque:1998kg, Braaten:2011sz}:
\begin{equation}
  \label{eq:h0}
  H_0(\Lambda)=0.879\, \frac{\sin(s_0 \ln(\Lambda/\Lambda_*)+\arctan(s_0))}
{\sin(s_0\ln(\Lambda/\Lambda_*)-\arctan(s_0))}~,
\end{equation}
where $s_0 = \pi/\ln\lambda= 1.00624$, and the multiplicative factor is a regulator-dependent number that can be obtained numerically~\cite{Braaten:2011sz}. $\Lambda_*$ is a three-body parameter that is determined by the LO renormalization condition. For example, it relates to $\kappa_*$ by $\Lambda_* = 0.548 \kappa_*$. As shown in both Fig.~\ref{pic:HLambda} and Eq.~\eqref{eq:h0}, $H_0$ is invariant under a discrete scale transformation $\Lambda \rightarrow \lambda \Lambda$. Therefore, three-body observables in the Efimov system display the discrete scaling symmetry due to the limit cycle behavior in the three-body Hamiltonian. 

\begin{figure}[ht]
\centerline{\includegraphics*[width=0.6\linewidth,angle=0,clip=true]{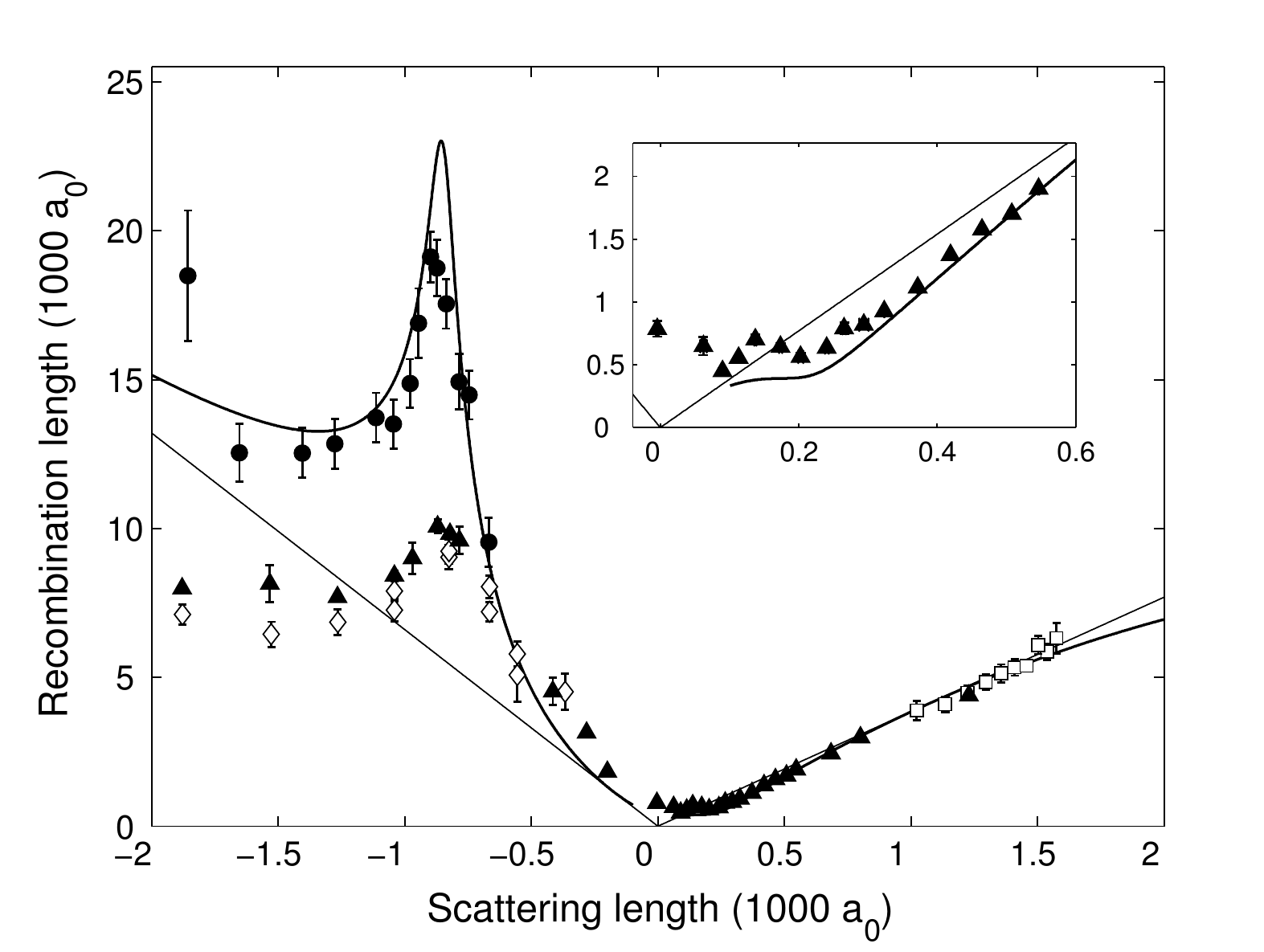}}
\caption{The recombination length $\rho_3=(2\sqrt{3}m\alpha)^{1/4}$ is measured in the ultracold $^{133}$Cs atomic gas as a function of $a$ and is compared with the EFT calculation. The solid curve represents the full LO EFT calculation including both shallow- and deep-dimer effects. The straight line indicates a lower limit for $a<0$ and an upper limit for $a>0$ from the EFT prediction. The figure is adapted from the original publication~\cite{Kraemer:2006} under copyright license no. 3714730785537.}
\label{pic:cs-rec}
\end{figure}

Another crucial Efimov feature related to $t_0$ is the three-body recombination rate measured in ultracold atomic gases. The recombination is a collision process in which three free
atoms collide and form a shallow dimer. The released energy is carried away kinetically by the third atom, which escapes the magnetic trap. The rate of the recombination process $\alpha$ is connected with the density of trapped atoms $n$ by $d n/dt=-3\alpha n^3$. In the zero-temperature approximation, $\alpha$ can be calculated using LO EFT by~\cite{Bedaque2000}  
\begin{equation}
\label{eq:alpha0-1}
\alpha=\frac{512\pi^2}{\sqrt{3}m}\left|t_0\left(0,\frac{2}{
\sqrt{3}a};0\right)\right|^2~.
\end{equation}
The existence of deep dimers in experimental setups opens additional recombination channels. A deep dimer, whose binding energy is of order $1/m \ell^2$ or larger, can occur regardless of the existence ($a>0$) or absence ($a<0$) of shallow dimers. If deep dimers exist, an additional recombination process occurs as the scattering between an atom and a deep dimer with large kinetic energy but small total energy. The deep-dimer effects on $\alpha$, which is beyond the EFT description for large-scattering-length physics, has been implemented by introducing a parameter $\eta_*$ encoding information of inelastic channels~\cite{Braaten:2003yc}. The resulting $\alpha_{\rm deep}$ is
\begin{subequations}
\begin{equation}
\alpha_{\rm deep} =
16.7 \, \left(1-e^{-4\eta_*} \right) \,\frac{\hbar a^4}{m}  \qquad (a>0); 
\end{equation}
\begin{equation}
\alpha_{\rm deep} = \frac{4590 \;\sinh(2\eta_*)}{\sin^2 [s_0 \ln (a/a_{-})] + \sinh^2 \eta_*}
\; \frac{\hbar a^4}{m} \qquad (a<0).
\end{equation}
\end{subequations}
As shown in Fig.~\ref{pic:cs-rec}, the EFT prediction in the recombination rate was successfully confirmed in the measurement of ultracold $^{133}$Cs atoms done at Innsbruck~\cite{Kraemer:2006}.

Deep dimers also open inelastic channels in the atom-dimer scattering, named the dimer relaxation. In this process, the deep dimer is formed by an atom scattering with a shallow dimer at low energies. The relaxation rate $\beta_{\rm relax}$ has also been calculated in the EFT framework as~\cite{Braaten:2003yc}
\begin{eqnarray}
 \beta_{\rm relax}=
\frac{20.3 \;\sinh (2\eta_*)}{\sin^2[s_0\ln(a/a_+)] 
 +  \sinh^2 \eta_*}\frac{\hbar a}{m}\,.
\label{beta-ad}
\end{eqnarray}
Experiments of the relaxation rate in $^{133}$Cs~\cite{knoop:2008} are in good agreement with the EFT predictions.

Although the ultracold atomic gases can be created at near-zero temperatures ranging from a few nK to a few $\mu$K~\cite{Kraemer:2006,knoop:2008,Zenesini2014,Huang2014,Gross:2009,Gross:2010,Pollack:2009,Zaccanti2009,Wild2012}, the finite-temperature effects make visible corrections to recombination and relaxation rates. Such effects have also been studied in the EFT framework~\cite{Braaten:2007pra,Braaten:2008pra}.


\subsection{Effective range corrections to Efimov physics}
\label{sec:1-3-atom}

Beyond the universal physics predicted in LO EFT, corrections from a finite effective range $r_0$ enter at higher orders in the EFT expansion. Such effects are visible in the experiments of ultracold atoms~\cite{Kraemer:2006,Pollack:2009,Gross:2009,Gross:2010,Zaccanti2009}, where the measured Efimov features, {\it i.e.} $a_*$, $a_+$ and $a_{-}$, display finite deviations from the universal relations predicted in Eq.~\ref{eq:a-kappa*}. The explanation of such discrepancies relies on including finite effective range into the EFT analysis of three-body systems.
Expanding observables in $r_0/a$, one can construct the EFT formalism~\cite{Hammer:2001gh,Bedaque:2002yg,Griesshammer:2004pe,Platter:2006ev,Ji:2010su, Ji:2011qg,Ji:2012nj} to analyze the range effects in three-body systems. 

\begin{figure}[ht]
\centerline{\includegraphics*[width=0.9\linewidth,angle=0,clip=true]{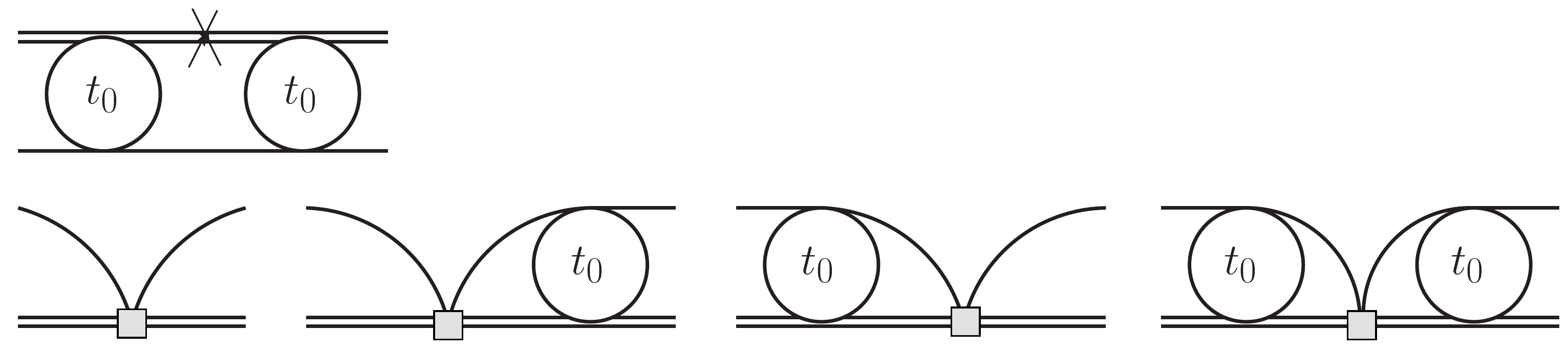}}
\caption{The diagrams contributing to the NLO atom-dimer scattering amplitude $t_1$ in perturbation theory: the crossed double and the square box denote respectively the NLO dimer propagator and the NLO three-body coupling.}
\label{pic:t1}
\end{figure}

The next-to-leading-order (NLO) correction to the atom-dimer scattering amplitude, $t_1$, is schematically represented by Fig.~\ref{pic:t1}, which corresponds to inserting the linear-in-$r_0$ pieces of the dimer propagator and three-body coupling, {\it i.e.} $\mathcal{D}^{(1)}(p)$ and $H_1(\Lambda)$, in the three-body system.

The NLO correction to the Efimov spectrum was calculated by Platter {\it et al.}~\cite{Platter:2008cx}. They demonstrated that the first-order-in-$r_0$ corrections to the trimer binding energies vanish in the unitary limit. The range corrections in systems with a fixed $a$ were studied in Refs.~\cite{Hammer:2001gh, Bedaque:2002yg,Griesshammer:2004pe} and applied to three-nucleon systems, where they concluded that renormalization at NLO does not depend on additional three-body observables. However, in cold atomic systems, the scattering length varies dramatically near the Feshbach resonance. Ji {\it et al.} developed a perturbative EFT formalism to study the range corrections to three-boson systems~\cite{Ji:2010su, Ji:2011qg}. They found that in systems with a variable scattering length, the NLO three-body coupling $H_1$ contains a ($r_0/a$)-dependent piece, which requires an additional three-body input for consistent renormalization at NLO. Refs.~\cite{Ji:2010su, Ji:2011qg} calculated the range corrections to the three-body recombination in $^7$Li atoms, which were measured at Bar-Ilan University~\cite{Gross:2009,Gross:2010} and Rice University~\cite{Pollack:2009}. The universal relations between Efimov features~\eqref{eq:a-kappa*} were reanalyzed in Ref.~\cite{Ji:2011qg} with the inclusion of linear-in-$r_0$ corrections and benchmarked with the $^7$Li experiments.

By analyzing the running of the three-body coupling in the renormalization group, Ji {\it et al.} recently revealed a pattern of the first-order range corrections to Efimov features~\cite{Ji2015}. The $r_0/a$ dependent piece in the three-body coupling, which contains a second three-body parameter, modifies the analytic expression of $H$ up to first-order in $r_0$ by
\begin{equation}
\label{eq:H1}
H(\Lambda) = H_0(\Lambda/\Lambda_*) + 0.351 \frac{ d H_0(\Lambda/\Lambda_*)}{d\ln \Lambda}\, \ln(\Lambda/\mu_0)  \frac{r_0}{a} +\cdots~,
\end{equation}
where $\mu_0$ encapsulates the additional three-body parameter besides $\Lambda_*$, and the ellipsis represents other linear-in-$r_0$ terms of $H$ that do not depend on $\mu_0$~\cite{Ji:2011qg}. The term proportional to
$\ln(\Lambda/\mu_0)$ in Eq.~\eqref{eq:H1} indicates a logarithmic breaking of the discrete scaling symmetry. It can be absorbed into the LO three-body coupling $H_0$ by replacing $\Lambda_*$ with a parameter $\tilde{\Lambda}_*(\mu_0,a)$ that runs logarithmically with the momentum scale at a rate proportional to $r_0/a$:
\begin{equation}
\label{eq:run-Lambda*}
\tilde{\Lambda}_*(\mu_0,a) \equiv
  (\mu_0/\Lambda_*)^{-0.351 r_0/a} \,  \Lambda_*~.
\end{equation}
The running three-body parameter yields new universal correlations among the range-corrected Efimov features, and modifies the LO universal correlations in Eq.~(\ref{eq:a-kappa*},\ref{eq:Buni},\ref{eq:ani}) by~\cite{Ji2015}
\begin{equation}
\label{eq:a--kappa*NLO}
a_{i,n} = \lambda^{n} \theta_i \kappa_*^{-1}
+ (J+\delta_i - n \sigma)  r_s,
\end{equation}
where $\delta_-=0$, $\delta_+=0.548$, $\delta_*=1.250$, and $\sigma=0.351\ln\lambda$. $J$ is tuned to reproduce the second three-body observable. This modified scaling symmetry showed good agreement with various potential-model-dependent calculations~\cite{Deltuva2012,Schmidt2012,Garrido2013}.

The Innsbruck group recently discovered that in experiments of different atomic species or at different Feshbach resonances the measured $a_-$ is always $-9$ times the van der Waals length $r_{vdw}$ within a $20\%$ accuracy~\cite{Berninger2011}. This correlation, named {\it van der Waals universality}, may originate from the underlying short-range interactions in various cold atomic systems. The origin of the van der Waals universality was studied in Refs.~\cite{Wang2012, Naidon2014, Blume2015, Horinouchi2015}. Using functional renormalization group, Horinouchi and Ueda recently described the van der Waals universality as the universality for the onset of the limit cycle~\cite{Horinouchi2015}, that is independent of short-range potentials. The insensitivity to underlying short-range physics suggests the connection between EFT and the van der Waals universality, which requires future investigations.

When the typical momentum scale in a few-body system is smaller, but within the same order of, the scale of the underlying short-range physics, range effects beyond the first order become important. This applies to few-nucleon systems where $r_0/|a| < 1/3$. In a system of three $^4$He atoms, although the excited-state trimer is shallowly bound as a signature of Efimov physics, an accurate description of the $^4$He trimer, especially for its ground state, requires analyzing corrections from higher-order range effects. 

The second-order range corrections to three-body observables can be determined by calculating the next-to-next-to-leading order (N$^2$LO) atom-dimer scattering amplitude in the EFT framework. 
Such N$^2$LO calculations for three-boson systems with a fixed $a$ were carried 
out by Bedaque {\it et al.}~\cite{Bedaque:2002yg}, 
by Grie{\ss}hammer~\cite{Griesshammer:2004pe} and by Platter and 
Phillips~\cite{Platter:2006ev}.
Although these works used a partial resummation formalism, they reached 
different conclusions on the running of the three-body coupling at this order.
Bedaque {\it et al.} showed that an additional, energy-dependent, three-body 
counterterm enters in N$^2$LO renormalization~\cite{Bedaque:2002yg}. This 
finding was later confirmed numerically by 
Grie{\ss}hammer~\cite{Griesshammer:2004pe}, who also extended this finding to 
three-nucleon systems in general settings~\cite{Griesshammer:2005ga}.
On the contrary, Platter and Phillips concluded that this additional piece vanishes in the limit $\Lambda \gg 1/r_0$~\cite{Platter:2006ev}. However, the partial resummation can arbitrarily include $\sim r_0^2$ (and beyond) range effects, which  cause complications to renormalization.

\begin{figure}[ht]
\centerline{\includegraphics*[width=0.95\linewidth,angle=0,clip=true]{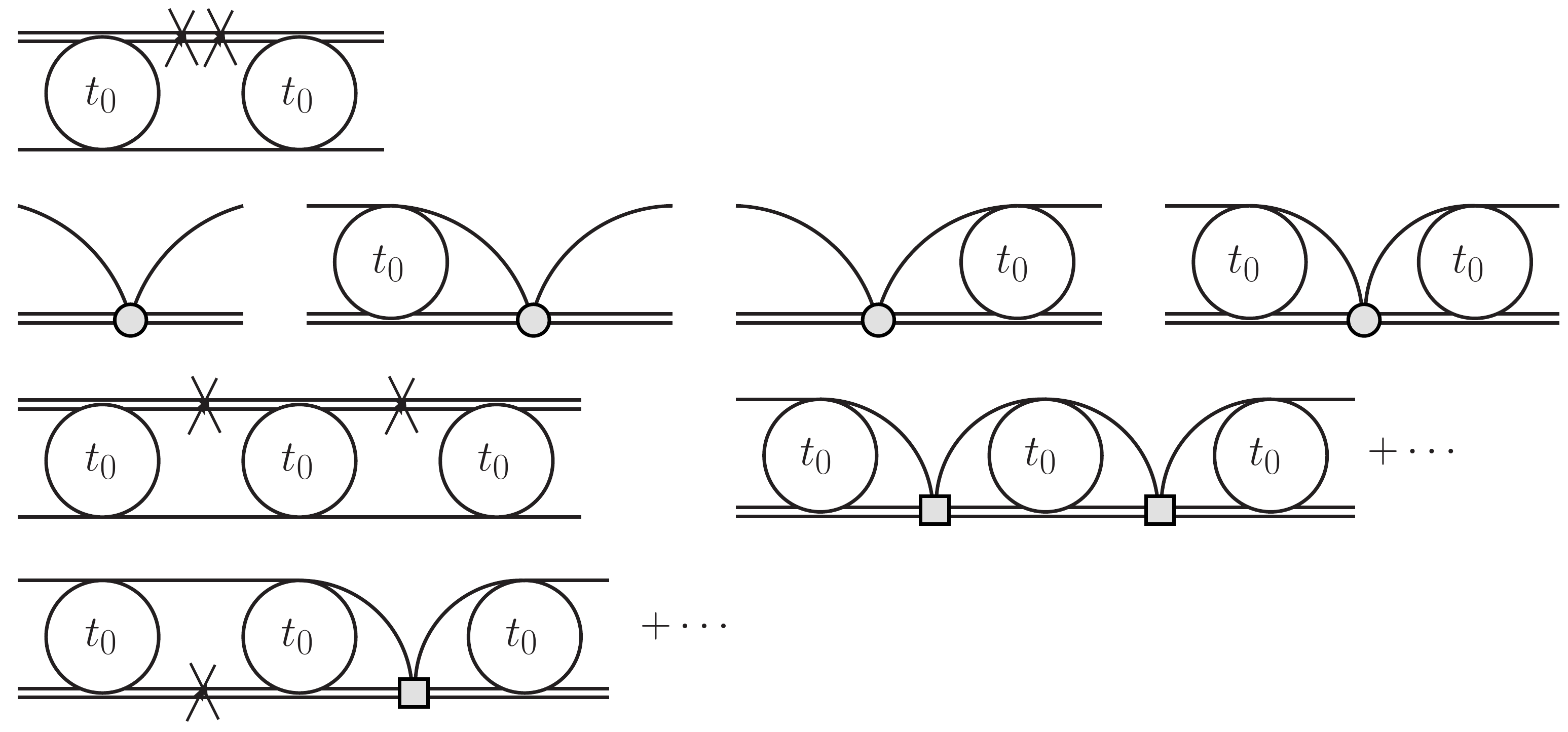}}
\caption{The diagrams contributing to the N$^2$LO atom-dimer scattering amplitude $t_2$ in second-order perturbation theory. The crossed (double crossed) double line denotes the NLO (N$^2$LO) dimer propagator, and the square (round) box represents the NLO (N$^2$LO) three-body counterterm.}
\label{pic:t2}
\end{figure}

To solve the controversy of higher-order three-body couplings raised in Refs.~\cite{Bedaque:2002yg,Griesshammer:2004pe,Platter:2006ev}, Ji and Phillips performed a rigorous perturbative three-body calculation up to N$^2$LO~\cite{Ji:2012nj}. The second-order range effects to the atom-dimer scattering amplitude were calculated using perturbation theory. As schematically shown in Fig.~\ref{pic:t2}, the contributions arise from the insertion of the N$^2$LO pieces of the dimer propagator and the three-body coupling, and from inserting twice the corresponding NLO terms.
Through regularizing the N$^2$LO atom-dimer amplitude, they verified the needs of the additional energy-dependent three-body coupling at N$^2$LO, which requires a second three-body input for proper renormalization. This indicates that the partial resummation is only consistent with a rigorous perturbative range expansion under the condition $\Lambda \sim 1/r_0$.

Ref.~\cite{Ji:2012nj} applied the rigorously perturbative range expansion to calculate the up-to-$r_0^2$ corrections to the trimer binding energies and atom-dimer scattering phase shifts in the system of three $^4$He atoms, and compared with calculations from realistic potentials~\cite{Roudnev2000, Roudnev2003} and the partially-resummed EFT~\cite{Platter:2006ev}. The results displayed a \mbox{$\sim$ 1-2\%} correction at NLO (\mbox{$\sim$ 1-3\permil} at N$^2$LO) for the excited-state $^4$He trimer binding energy and atom-dimer scattering length, and a \mbox{$\sim$ 10\%} correction at NLO (\mbox{$\sim$ 20\%} at N$^2$LO) for the ground-state $^4$He trimer binding energy.

The perturbative $r_0/a$ expansion formalism has also been developed by Vanasse {\it et al.}~\cite{Vanasse:2013sda,Vanasse:2014kxa}, in the pionless EFT framework, to investigate range effects in few-nucleon bound-state and scattering problems.


\subsection{Triatomic systems beyond identical bosons}

Besides in systems of ultracold bosonic atoms, the Efimov scenario can also be observed in a quantum degenerate gas of fermionic atoms. The Efimov trimers have been measured in the $^6$Li atomic mixture of three degenerate hyperfine states. In this system, the $^6$Li dimers have three different scattering lengths which vary respectively near their own Feshbach resonances. Due to the overlap of these resonance regimes, three scattering lengths can be tuned simultaneously through an external magnetic field, creating a three-hyperfine-component atomic mixture.

Several experiments have measured the recombination features of the three-component $^6$Li atoms associated with the ground-state Efimov trimer~\cite{Ottenstein2008,Huckans2009} and the excited-state trimer~\cite{Williams:2009}. Shortly after these experiments, Nakajima {\it et al.}~\cite{Nakajima2010} observed an enhanced resonance of a dimer relaxation process in $^6$Li due to the existence of Efimov trimers. The binding energy of the $^6$Li trimer was determined by the radio-frequency association method~\cite{Nakajima2011}.

Using contact EFT approaches, Braaten {\it et al.} calculated various aspects of the Efimov physics in the three-spin mixture of $^6$Li atoms~\cite{Braaten:2008wd,BHKP2010}. With three scattering lengths at different but large values in this system, they constructed the coupled-channel STM equations to calculate the spectrum of Efimov trimers, the three-body recombination rate, and the dimer relaxation rate near the resonance when the Efimov trimer crosses the atom-dimer scattering threshold. The universal properties in $^6$Li atoms, predicted by the EFT calculations~\cite{Braaten:2008wd,BHKP2010} in the $r_0\rightarrow0$ limit, displayed good agreements with experimental results~\cite{Ottenstein2008,Huckans2009,Williams:2009,Nakajima2010,Nakajima2011}.

Another area to observe Efimov scenario is in heteronuclear systems, which are a mixture of two ultracold atomic species near an interspecies Feshbach resonance. 
The heteronuclear system opens new possibilities to study the discrete scaling symmetry, which is a critical feature at the heart of Efimov physics. The scaling constant $\lambda=22.694$ in systems of three identical bosons yields a large energy gap between the ground- and excited-state Efimov trimers. This gap makes the experimental observation of an excited Efimov state, which is already near the saturation regime due to thermal motion, a very challenging task. This obstacle can be potentially circumvented in heteronuclear systems.    
The effective field theory calculation by Helfrich {\it et al.} indicated the existence of Efimov physics in a heteronuclear trimer containing a mixture of two atomic species, which has a large interspecies scattering length and a small intraspecies scattering length~\cite{Helfrich:2010yr}. Their studies found that the scaling factor is much larger than $22.694$ for a heteronuclear trimer containing two species with comparable masses, but much smaller in a trimer consisting of one light and two heavy atoms. The later case makes the observation of several Efimov states connected by a small geometric scaling possible. 

Several experiments have investigated the recombination and relaxation in a Bose-Bose mixture of $^{41}$K-$^{87}$Rb atoms~\cite{Barontini:2009} and a Fermi-Bose mixture of $^{40}$K-$^{87}$Rb~\cite{Bloom:2013}. Although the scaling constant is large in these K-Rb trimers, {\it i.e.} 131 for $^{41}$K-$^{87}$Rb-$^{87}$Rb, 348000 for $^{87}$Rb-$^{41}$K-$^{41}$K, and 123 for $^{40}$K-$^{87}$Rb-$^{87}$Rb predicted from EFT~\cite{Helfrich:2010yr},  the observed resonance features, associated with ground-state trimers of the K-Rb mixture, showed reasonably good agreement with EFT calculations in the zero-range limit. The theory-experiment deviation is due to finite range effects and finite temperature effects.

Recent experiments in the $^{6}$Li-$^{133}$Cs mixture, that has an extreme mass imbalance, have observed the discrete scaling symmetry~\cite{Pires:2014,Tung:2014,Ulmanis:2015,Ulmanis:2015nw}. By measuring the three-body loss resonance at negative scattering lengths near a $^{6}$Li-$^{133}$Cs Feshbach resonance,  Pires {\it et al.} discovered a ratio of $5.8$ between the two $a_{-}$'s associated with the ground- and excited-state $^{6}$Li-$^{133}$Cs-$^{133}$Cs~\cite{Pires:2014}. A later work by Tung {\it et al.} observed a geometric scaling factor $4.9$ among three consecutive $^{6}$Li-$^{133}$Cs-$^{133}$Cs Efimov trimers~\cite{Tung:2014}. These measured scaling constants are consistent with a universal zero-range theory calculation, finding $\lambda=4.88$~\cite{Petrov2015}. This work by Petrov and Werner also implemented effects from finite temperatures and Cs-Cs interactions into the zero-range theory. Their results displayed good agreement with the measured three-body recombination in the $^{6}$Li-$^{133}$Cs mixture.


\section{Universal physics in halo nuclear systems}
\label{sec:1-4-halo}

\subsection{Constructing effective field theory for halo nuclei}

Halo nuclei, either neutron rich or proton rich, are found to have large nuclear charge/matter radii and shallow binding energies. For the scale separation in halo nuclei, 
the short-range scale refers to the size of the nuclear core, $R_{\rm core}$, and the large-distance scale is characterized by the average distance from a valence nucleon to the core, $R_{\rm halo}$, satisfying $R_{\rm core} \ll R_{\rm halo}$. One can study physics in halo nuclei by treating the core as an inert point-like particle and work in cluster degrees of freedom consisting of valence nucleons and a core. In effective field theory for halo nuclei, the contact interactions among valence nucleons and the core are constructed based on the expansion of $R_{\rm core}/R_{\rm halo}$. The dynamic core-excitation effects and the effects of valence- and core-nucleon anti-symmetrization are embedded in contact interactions.

The EFT for halo nuclei was first constructed by Bertulani {\it et al.}~\cite{Bertulani:2002sz} and Bedaque {\it et al.}~\cite{Bedaque:2003wa}. They studied $^5$He as a neutron-$\alpha$ two-body system interacting in a p-wave resonance, where different power-counting approaches based on the $R_{\rm core}/R_{\rm halo}$ expansion are employed in these two works.
By implementing the long-range Coulomb potential, in addition to the short-range interaction, into the EFT framework, theoretical investigations have been extended to $^8$Be as an $\alpha$-$\alpha$ scattering resonance~\cite{Higa:2008dn}, the proton-$\alpha$ scattering~\cite{Higa:2008fbs,Higa:2011fbs}, and $^{17}$F as a one-proton-halo~\cite{Ryberg:2014}.

EFT studies of halo nuclei with one valence nucleon have also been applied to calculate electric-dipole strengths in the photo-dissociation of $^{11}$Be~\cite{Hammer201117} and $^{19}$C~\cite{Acharya2013103}.
Recently, the EFT method for halo nuclei have been applied actively to describe the radiative neutron-captures on $^7$Li~\cite{Rupak2011,Fernando2012,Zhang:2014n} and $^{14}$C~\cite{Rupak:2012prc}, and the radiative proton-captures on $^7$Be~\cite{Zhang:2014p,Ryberg:2014epja,Zhang:2015ajn} and $^{16}$O~\cite{Ryberg:2014}.

The halo physics also exists in the three-body sector, which typically involves a halo nucleus with two valence neutrons ($2n$ halo). The neutron-core ($n$-$c$) interaction is parameterized in the EFT framework through the effective range expansion of the partial-wave decomposed t-matrix. If the $n$-$c$ t-matrix is dominated at low-energies by a resonance in a particular partial wave $L$ ($L=\textrm{s},\textrm{p},\textrm{d},\cdots$), we refer to such nucleus as an $L$-wave halo.

\begin{figure}[ht]
\centerline{\includegraphics*[width=0.95\linewidth,angle=0,clip=true]{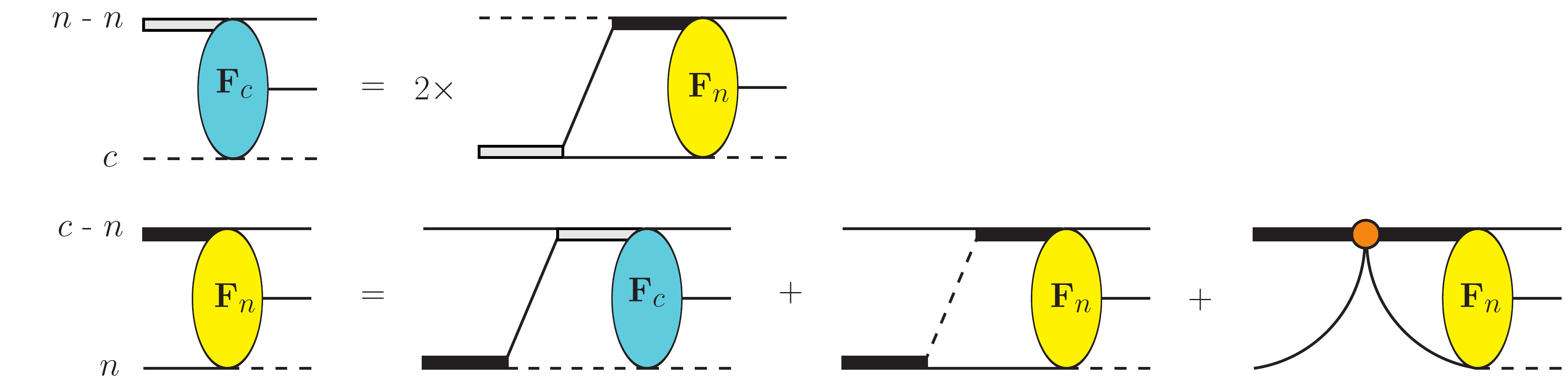}}
\caption{The bound-state scattering amplitudes $F_c$ and $F_n$ in coupled integral equations. The solid and dashed lines are respectively the $n$ and $c$ propagators, the double line and thick line denote respectively the $nn$ and $nc$ dimeron propagator, and the round box represents the $nnc$ three-body force.}
\label{pic:halo-stm}
\end{figure}

The EFT formalism for a $2n$ halo is similar to that for a heteronuclear trimer of one heavy bosonic and two light fermionic atoms. However, differently from the heteronuclear system, where only the interspecies scattering length is large, the interaction between the two valence neutrons, dominated by a low-energy s-wave virtual state, also has a large scattering length. Therefore, the STM equation for an $L$-wave $2n$ halo requires both the $n$-$n$ s-wave and $n$-$c$ $L$-wave interactions, represented by their own {\it dimeron} operators. The bound-state problem is to solve the scattering amplitudes of $c$-($nn$) ($F_c$) and $n$-($nc$) ($F_n$)  in coupled homogeneous integral equations, as illustrated schematically in Fig.~\ref{pic:halo-stm}.  A $n$-$n$-$c$ three-body counterterm is also needed for renormalization. In the following, I focus on the EFT achievements in calculating s- and p-wave $2n$ halo nuclei.

\subsection{Three-body s-wave halo nuclei} 

In an s-wave $2n$ halo nucleus, the neutron-core interaction is dominated by a low-energy s-wave bound (virtual) state with a large positive (negative) scattering length. Due to its similarity to a heteronuclear trimer, the s-wave $2n$ halo engenders tremendous theoretical and experimental interest in unveiling Efimov effects and the discrete scaling symmetry in halo nuclei.

\begin{figure}[ht]
\centerline{\includegraphics*[width=0.65\linewidth,angle=0,clip=true]{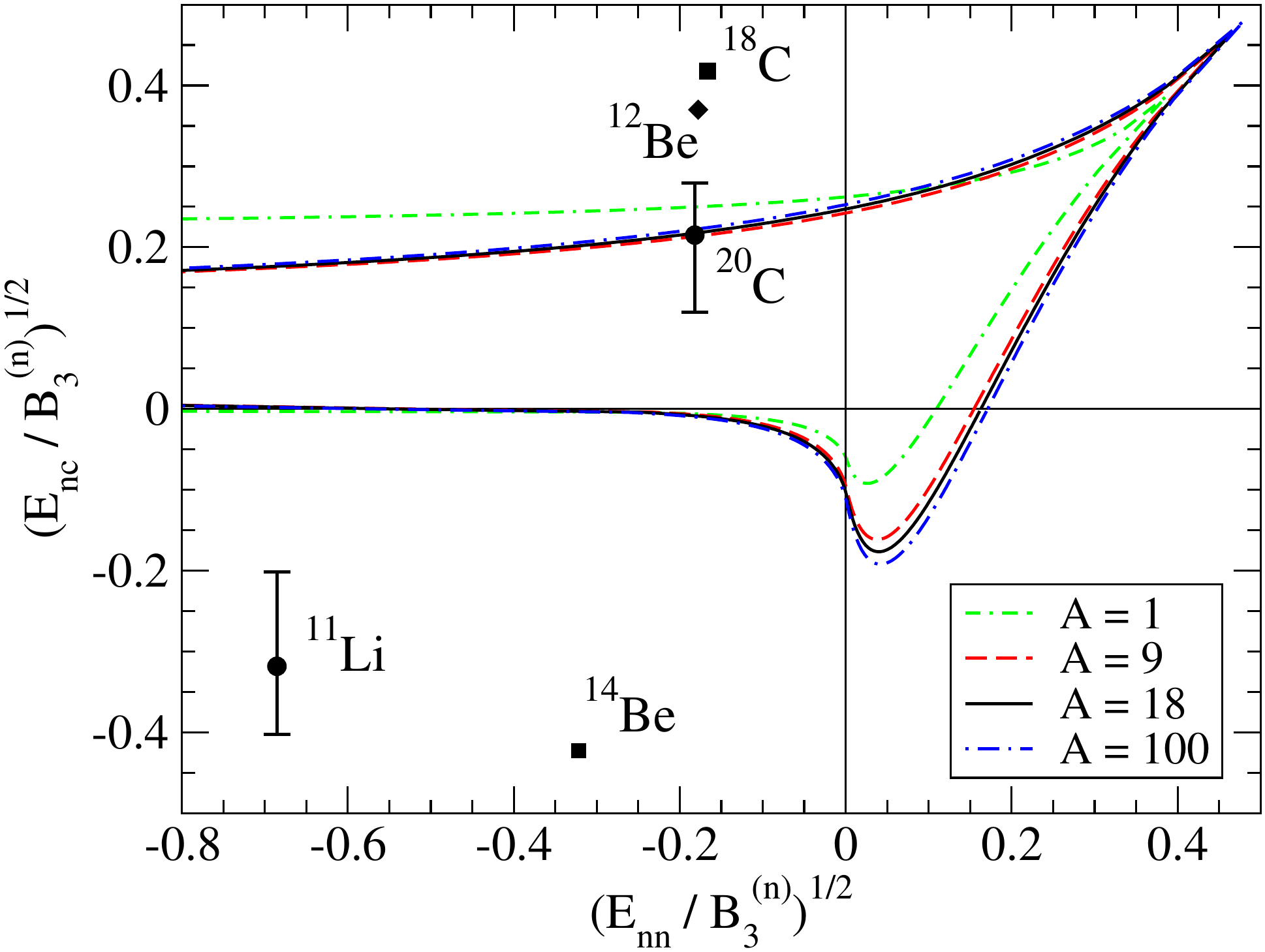}}
\caption{The contour plot in $ E_{nc}/B_3$ versus $E_{nn}/B_3$ for the ground-state $2n$ halos at mass numbers $A$ of the cores. The excited Efimov state is at threshold along the contour curves.
The figure is adapted from the original publication~\cite{Canham:2008jd} under copyright license no. 3718851192616.}
\label{pic:halo-efimov}
\end{figure}

Canham and Hammer have performed EFT calculations of binding energies, root-mean-square (rms) matter radii, and three-body configuration distributions of several s-wave $2n$ halos~\cite{Canham:2008jd}, with finite range effects included in a following paper~\cite{Canham:2009xg}.
Working in the zero-range limit, they also explored the possibilities of finding an excited Efimov state in different halo nuclei, by taking inputs from the experimental values of the $2n$ separation energy $B_3$, neutron-core energy $E_{nc}$ ($E_{nc}>0$ for a bound state and $E_{nc}<0$ for a virtual state), and the neutron-neutron energy $E_{nn}$ of a given nucleus.
They constructed a boundary region constrained by the contour plot of the ratios $ E_{nc}/B_3$ versus $E_{nn}/B_3$. As shown in Fig.~\ref{pic:halo-efimov}, the contour curve converges to a universal region by increasing the mass number of the core $A$. If there exists an excited Efimov state below the $nnc$ breakup threshold, the measured ground-state properties of a $2n$ halo yield a point inside the boundary region. This analysis was applied to $^{11}$Li, $^{18,20}$C and $^{12,14}$Be as halo candidates based on experimental data from  the ``Nuclear Data Evaluation Project'' of TUNL~\cite{TUNL}. They showed that the only possible candidate for an excited Efimov state is $^{20}$C, mainly due to the large uncertainty of the experimental $E_{nc}$ of $^{20}$C, {\it i.e.} $162\pm112$ keV~\cite{TUNL}. Similar searches for Efimov states were also done by Frederico, Tomio and collaborators in a zero-range three-body model~\cite{Amorim:1997mq,Yamashita:2007ej}. However, newer experimental data, from the Coulomb dissociation of $^{19}$C ($E_{nc}=530\pm130$ keV~\cite{Nakamura:1999prl}) and the 2012 atomic mass evaluation (AME2012) ($E_{nc}=580\pm90$ keV~\cite{Wang:2012ame}), excluded an excited Efimov state in $^{20}$C.

The most neutron-rich carbon isotope $^{22}$C is another halo candidate.
The measurement of the reaction cross section of $^{22}\text{C}$ on a hydrogen target~\cite{Tanaka:2010zza} determined a $^{22}\text{C}$ rms matter radius $\sqrt{\langle r_m^2\rangle} =5.4\pm 0.9$~fm. This measurement also suggests that the $^{22}\text{C}$ ground state has an s-wave $2n$ halo configuration, where the $^{21}\text{C}$ subsystem is known to be unbound~\cite{Wang:2012ame}. This halo picture was confirmed by a neutron removal reaction measurement on neutron-rich carbon isotopes~\cite{Kobayashi:2011mm}.
Except $\sqrt{\langle r_m^2\rangle}$, other properties of $^{22}\text{C}$ were measured with very large uncertainties. From AME2012~\cite{Wang:2012ame}, the $n$-$^{20}\text{C}$ interaction has a continuum energy of $E_{nc}=-10\pm470$ keV, while the three-body binding energy of $^{22}\text{C}$ is $B=110\pm60$ keV.   

\begin{figure}[ht]
\centerline{\includegraphics*[width=0.65\linewidth,angle=0,clip=true]{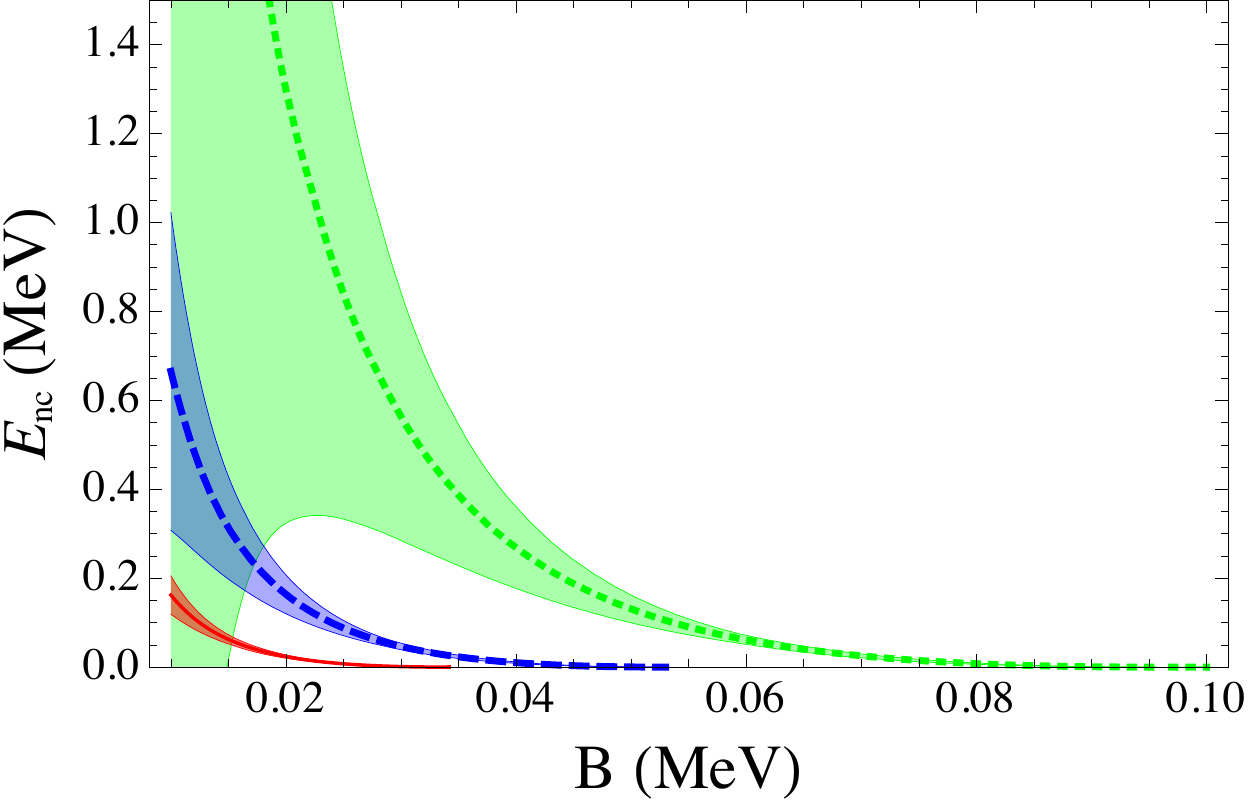}}
\caption{The correlation curve of $B$ versus $E_{nc}$ in $^{22}\text{C}$ with fixed values of the matter radius at $\sqrt{\langle r^2 \rangle} $~=~5.4~fm (blue, dashed), 6.3~fm (red,solid), and 4.5~fm (green, dotted). The shaded bands indicate the theoretical errors based on estimates of higher-order EFT corrections.}
\label{pic:22C-contour}
\end{figure}

Acharya {\it et al.}~\cite{Acharya:2013aea} applied the EFT analysis and predicted the universal correlations among $\sqrt{\langle r_m^2\rangle}$, $E_{nc}$ and $B$ in $^{22}$C. As shown in Fig.~\ref{pic:22C-contour}, three sets of $B$ vs $E_{nc}$ correlation are plotted in Ref.~\cite{Acharya:2013aea} by fixing the experimental values of $\sqrt{\langle r_m^2\rangle}$ within a $\pm 1\sigma$ uncertainty, {\it i.e}, 4.5~fm, 5.4~fm and 6.3~fm, along with the theoretical error bands from estimating the $R_{\rm core}/R_{halo}$ corrections. The correlation plot indicates that the $\sqrt{\langle r_m^2\rangle}$ data~\cite{Tanaka:2010zza} sets an upper limit of 100 keV on the $2n$ separation energy $B$ of $^{22}\text{C}$, regardless of the poorly determined $E_{nc}$. This upper bound on $B$ is consistent with the AME2012 data. It is also $20\%$ lower than the calculation in a zero-range three-body model~\cite{Yamashita201190}, and significantly lower than other model-dependent calculations~\cite{Horiuchi:2006prc,Fortune:2012prc}. Based on the $E_{nc}$ and $B$ correlation, Acharya {\it et al.}~\cite{Acharya:2013aea} explored the possibility for an excited Efimov state of $^{22}$C, whose existence only occurs if the $n$-$^{20}$C continuum energy is of the order of a few keV.

Based on the EFT approach, Hagen {\it et al.} developed a formalism to calculate the charge form factor $F(q)$ of s-wave $2n$ halos~\cite{Hagen:2013epja}. The low-momentum expansion of the calculated charge form factor is related to the charge radius of a halo nucleus $\sqrt{\langle r_c\rangle^2}$ by
\begin{equation}
\label{eq:rc-ff}
F(q) = 1- \langle r_c\rangle^2 q^2/6 +\cdots.
\end{equation}
$\sqrt{\langle r_c\rangle^2}$ shows universal correlations with other observables in $2n$ halos, such as $B$, $E_{nc}$, and $\sqrt{\langle r_m^2\rangle}$. These correlations were applied in Ref.~\cite{Hagen:2013prl} to 
investigate $^{62}$Ca as the heaviest $2n$ halo nucleus, where the information on the n-core interaction was obtained from an {\it ab initio} coupled-cluster calculation of the $n$--$^{60}$Ca s-wave scattering phase shift. This combination of EFT and {\it ab initio} methods predicted Efimov physics in $^{62}$Ca.

The dipole response functions of $^{11}$Li and $^{22}$C, which are associated with their Coulomb dissociation cross sections, have been recently calculated as s-wave $2n$ halo systems using a leading-order halo EFT~\cite{Acharya:2015gpa}. In the case of $^{11}$Li, the calculation showed an agreement with the experimentally measured Coulomb dissociation~\cite{Nakamura:2006prl} at low transition energies, within the error bars arising from corrections beyond the leading order of the halo EFT.


\subsection{Three-body systems of p-wave halo nuclei}

The p-wave two-body phase shift can be expanded in the low-energy limit by
\begin{equation}
\label{eq:kcot1}
k^3\cot\delta_1 = -\frac{1}{a_1} + \frac{1}{2}r_1 k^2 +\cdots~,
\end{equation}
where $a_1$ is the scattering volume, and the p-wave effective ``range'' $r_1$ has the dimension of momentum. 
Recent studies~\cite{Braaten:2011vf,Nishida:2011np} indicated that in a three-body system with resonant pairwise p-wave interactions, the discrete scaling symmetry cannot be realized in the limit $|a_1|\rightarrow\infty$ and $r_1\rightarrow 0$ because this p-wave unitary limit is not physical: it inevitably yields a two-body state with a negative probability density and therefore violates the causality condition.

Despite the absence of discrete scaling symmetry in p-wave-pairwise three-body systems, low-energy observables in such systems can still display interesting universal features. One example is $^6$He, which is a p-wave $2n$ halo nucleus with the neutron-core ($n$-$\alpha$) interaction dominated by a $^2$P$_{3/2}$ resonance. The existence of a three-body bound state in $^6$He shows the evidence of universal physics controlled by low-energy dynamics in higher partial waves.

\begin{figure}[ht]
\centerline{\includegraphics*[width=0.6\linewidth,angle=0,clip=true]{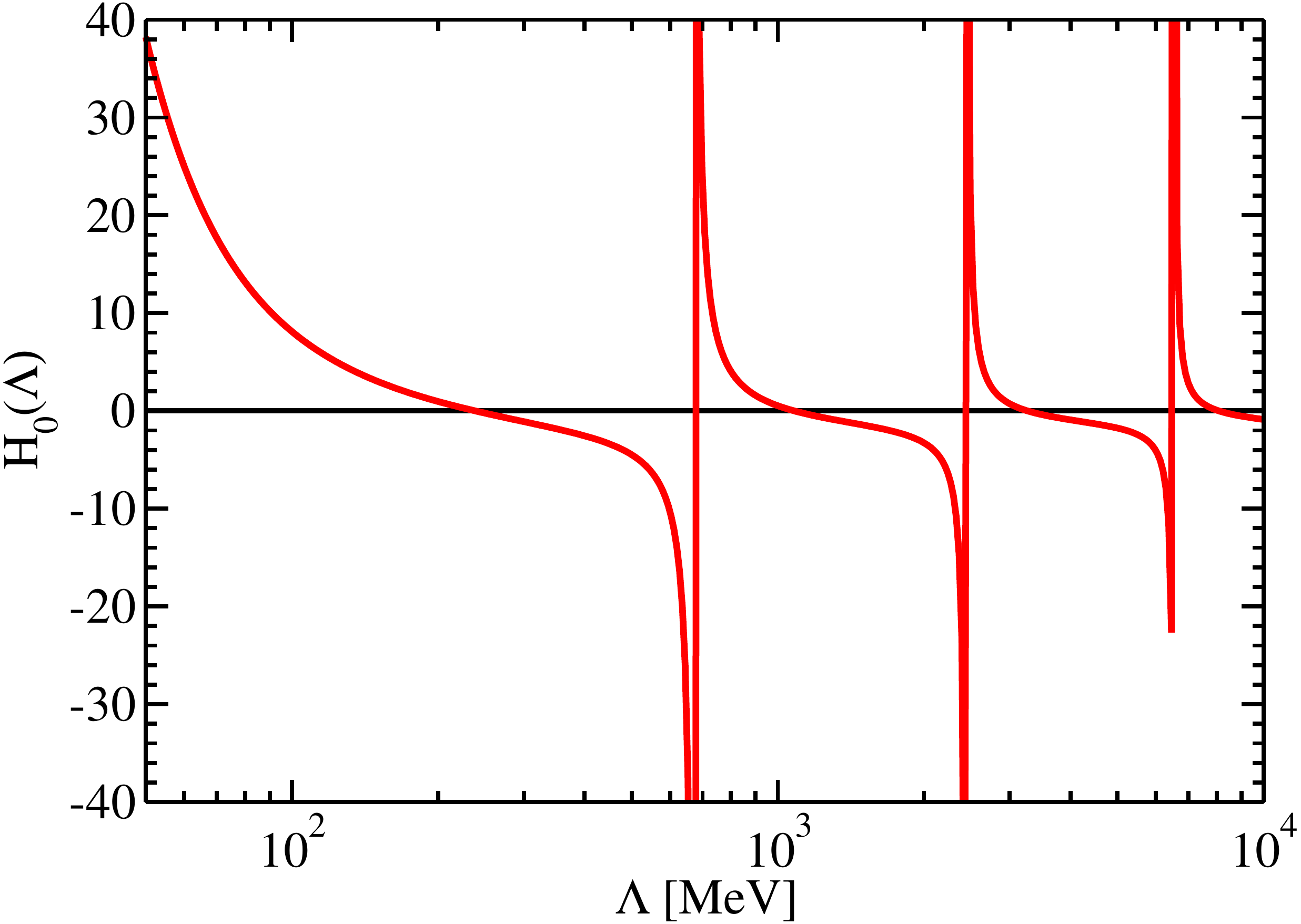}}
\caption{The $nn\alpha$-counterterm parameter $H_0$ as a function of the cutoff $\Lambda$. $H_0$ is tuned to reproduce $B_3=0.975$ MeV at different values of $\Lambda$.}
\label{pic:6he-3bf}
\end{figure}

Ji {\it et al.} described the $^6$He ground state as a $nn\alpha$ system in the halo EFT framework~\cite{Ji:2014wta}. The $n\alpha$ ${}^2$P$_{3/2}$ interaction was expanded by employing a ``narrow resonance'' power counting, {\it i.e.} $a_1 \sim R_{\rm core} R_{\rm halo}^2$ and $r_1 \sim 1/R_{\rm core}$, developed by Bedaque {\it et al.}~\cite{Bedaque:2003wa}. This power counting avoids the expansion near the unphysical unitary limit but preserves the low-energy p-wave resonance.
To obtain a renormalized $2n$ separation energy $B$ in the integral equations (Fig.~\ref{pic:halo-stm}) of the $nn\alpha$ system, Ji {\it et al.} introduced a p-wave $nn\alpha$ counterterm. By reproducing the experimental value $B=0.975$ MeV, they analyzed the running of the three-body coupling $H_0$ as a function of the regulation cutoff $\Lambda\sim1/R_{\rm core}$. As shown in Fig.~\ref{pic:6he-3bf}, $H_0$ exhibits a log-periodic behavior with decreasing periods when $\Lambda$ increases. This different behavior from $H_0$ in three-boson systems (Fig.~\ref{pic:HLambda}) indicates the breaking of discrete scaling invariance by the presence of the p-wave pairwise interaction. The $^6$He ground state was also calculated by Rotureau and van Kolck~\cite{Rotureau:2012yu} by combining EFT potentials with the Gamow-shell-basis-expansion few-body techniques.

If the three-body behavior in $^6$He is universal, the EFT analysis can also be applied to investigate other p-wave $2n$ halo nuclei. One example is $^{11}$Li, which is another Borromean $2n$ halo system. The measurement of the $2n$ transfer reaction, $^1$H($^{11}$Li,$^9$Li)$^3$H, implied that both the s- and p-wave components of $n-^{9}$Li interactions play important roles in forming the $^{11}$Li ground state~\cite{Tanihata:2008vw}.
Therefore, the binding and structural properties of $^{11}$Li may display unique three-body features that depend on both s- and p-wave two-body and three-body interactions.

\section{Conclusion}

Three-body systems display universal features at large distance scales, when the range of the pairwise interaction is much shorter than the two-body scattering length. The universal three-body physics exists in systems spanning over very different energy scales, including few-nucleon, cold atomic, and halo nuclear systems. Utilizing the scale separation between the short interaction range $\ell$ and the large scattering length $a$, low-energy behaviors in three-body systems can be investigated in the effective field theory framework by constructing two- and three-body contact interactions based on an $\ell/a$ expansion.

In this review, we have discussed the effective field theory studies of three-body universal physics with an emphasis on cold atomic and halo nuclear systems. Universal correlations among the Efimov features, related to critical phenomena of the three-body recombination and dimer relaxation in ultracold bosonic atomic gases, were predicted by EFT in the zero-range limit. The discrete scale invariance was explained, in renormalization group, through the running of the three-body coupling built in EFT. Finite-range corrections to Efimov physics, which breaks the discrete scaling symmetry, are also explained in the EFT framework as higher-order effects in the $\ell/a$ expansion. The investigation of range effects is applied to recombination features in ultracold atoms and the binding and scattering properties of $^4$He atomic trimers. Studies of Efimov physics are also extended to the inclusion of deep-dimer effects in cold atoms, fermionic atoms with spin mixtures, and heteronuclear atomic mixtures. 

In halo nuclei, the size of the nuclear core is much smaller than the size of the halo. This scale separation provides three-body universal physics in halo nuclei with two valence neutrons. A number of two-neutron halo nuclei with either s- or p-wave pairwise interactions are investigated using halo EFT. Studies of universal correlations among the halo spectrum at low-energies, structural distributions, charge and matter-radii, and electro- and photo-induced reactions are discussed in this review. These universal correlations are also combined with experimental constraints to explore the possibility of finding an excited Efimov state in halo nuclei.

\section{Acknowledgments}
The author is grateful to Lucas Platter and Wolfram Weise for valuable comments on the manuscript.
This research was supported in part by the Natural Sciences and Engineering Research Council (NSERC) and the National Research Council Canada.

\end{document}